\documentclass[aps,prl,reprint,superscriptaddress,showpacs,columnline]{revtex4-2}
\usepackage{hyperref}
\usepackage{graphicx}
\usepackage{dcolumn}
\usepackage{bm}
\usepackage{amsmath}
\usepackage{newtxtext,newtxmath}
\usepackage[T1]{fontenc}
\usepackage{ae,aecompl}
\usepackage{hyperref}
\usepackage{footnote}
\usepackage{graphicx}	
\usepackage{dblfloatfix}
\usepackage{amsmath}	
\usepackage{threeparttable}
\usepackage{subcaption}
\usepackage{caption}
\usepackage{color}
\usepackage[dvipsnames]{xcolor}
\usepackage[normalem]{ulem}
\usepackage{booktabs}
\usepackage{multirow}
\usepackage{mathtools}
\newcommand{\gray}{gamma-ray~}
\newcommand{\grays}{gamma-rays~}

\begin{document}

\title{Precise measurement of pion-bump structure using future MeV gamma-ray detectors}

\author{Jiahao Liu}
\affiliation{Deep Space Exploration Laboratory/School of Physical Sciences, University of Science and Technology of China, Hefei 230026, China}
\affiliation{CAS Key Laboratory for Research in Galaxies and Cosmology, Department of Astronomy, School of Physical Sciences,\\
University of Science and Technology of China, Hefei, Anhui 230026, China}
\affiliation{School of Astronomy and Space Science, University of Science and Technology of China, Hefei, Anhui 230026, China}

\author{Bing Liu}
\affiliation{Deep Space Exploration Laboratory/School of Physical Sciences, University of Science and Technology of China, Hefei 230026, China}
\affiliation{CAS Key Laboratory for Research in Galaxies and Cosmology, Department of Astronomy, School of Physical Sciences,\\
University of Science and Technology of China, Hefei, Anhui 230026, China}
\affiliation{School of Astronomy and Space Science, University of Science and Technology of China, Hefei, Anhui 230026, China}

\author{Ruizhi Yang}
\email{yangrz@ustc.edu.cn}
\affiliation{Deep Space Exploration Laboratory/School of Physical Sciences, University of Science and Technology of China, Hefei 230026, China}
\affiliation{CAS Key Laboratory for Research in Galaxies and Cosmology, Department of Astronomy, School of Physical Sciences,\\
University of Science and Technology of China, Hefei, Anhui 230026, China}
\affiliation{School of Astronomy and Space Science, University of Science and Technology of China, Hefei, Anhui 230026, China}

\begin{abstract}

The pion-bump structure in the gamma-ray spectrum is a direct proof for the hadronic origin of the gamma rays, and thus the decisive evidence for the acceleration of hadronic cosmic rays in astrophysical objects. However, the identification of such a spectral feature is limited by the resolution and energy coverage of current gamma-ray instruments. Furthermore, there are unavoidable bremsstrahlung emissions from secondary and primary electrons, which may dominate the gamma-ray emission below the pion-bump. Thus, the study of this gamma-ray emission component can provide unique information on the acceleration and confinement of high-energy particles.  In this paper, we studied the predicted gamma-ray spectrum assuming both hadronic or leptonic origin in mid-aged supernova remnants W44, we discuss the detection potential of future MeV missions on these emissions and possible implications.

\end{abstract}
\maketitle
\section{Introduction}
\label{sec:intro}
The origin of cosmic rays (CRs) is one of the most fundamental questions in modern astrophysics. Supernova remnants (SNRs) are regarded as the most promising acceleration sites of CRs \citep[see, e.g., ][]{blasi13}. Thus, they are also prime targets in \gray emissions in both GeV and TeV energy range \citep{fermi_snr, hgps}. The origin of high-energy \grays from SNRs is either from the inverse Compton (IC) scattering of relativistic electrons or from the pion-decay process in the inelastic scatterings of CR protons with ambient gas, while the latter is regarded as a strong proof that SNRs can accelerate CRs.  However, the discrimination of leptonic and hadronic scenarios is not easy, since in the energy range above GeV both mechanisms can produce the power-law spectra as observed. One distinct spectral feature of the hadronic scenario is the so-called "pion bump", which is a sharp low-energy spectral cutoff due to the kinematics of the neutral pion decay process. 

AGILE and Fermi-LAT collaborations reported the discovery of pion bumps in mid-aged SNR W44 and IC~443, which is regarded as decisive evidence that SNRs do accelerate hadronic CRs \citep{Agile2011W44,fermi_13}. However, later studies have argued that the observed feature can also be attributed to the bremsstrahlung of electrons \citep{Peron2020}, especially taking into account that the electrons in the ISM also reveal a low-energy break at several $\rm GeV$ \citep{strong11}. The observed low-energy break in both scenarios is most evident in the energy range between 10--100~MeV.  Due to the limited sensitivity of current \gray instruments in this energy range, it is rather difficult to distinguish the two scenarios. 
Furthermore, even in the hadronic scenario, in which the GeV \gray emission are dominated by the pion-decay process, there are inevitably accompanied primary electrons accelerated at the same site as well as the secondary electrons produced by the decay of charged pions in the inelastic collision of CR protons with ambient gas. These electrons can produce \grays in the same environment via bremsstrahlung and may dominate the \gray emissions below the pion bump. Thus the measurement of the \gray emissions below the pion bump can be used to study the the physical conditions of the accelerators, such as the e/p ratio and particle confinement in the vicinity \citep{Yang2018}.

In this paper, we will study the possible advance in the detection of the pion bump in one of the most suitable sites for such kind of study, the mid-age SNR W44, using the next-generation MeV \gray detectors.  The paper is organized as follows. In Sec.2, we briefly introduce the SNR W44 and the current \gray observations,  we then calculate the \gray flux in the MeV band assuming either the \grays are produced in pion-decay process (hadronic scenario) or bremsstrahlung (leptonic scenario).  Next, in Sec.3, we calculate the \gray emissions from primary/secondary electrons in the context of hadronic scenarios. Then in Sec.4, we estimate the detection ability of next-generation MeV detectors in this energy range.  Finally, we discuss the possible implications of our calculation in Sec.5.

\section{Gamma-rays from SNR W44}

\label{sec:w44}

W44 (G034.7-00.4) is an old mixed-morphology SNR that has a radio continuum shell filled with thermal X-ray emission from the shock-heated gas. A pulsar wind nebula associated with pulsar PSR B1853+01 is also embedded within the SNR \citep{1996ApJ...464L.161H}. { Although they could be potential MeV emitters, the X-ray flux is less than 1$\times10^{-14}$~erg~s$^{-1}$~cm$^{-2}$ as mentioned in \cite{1996frail}. If we assume the spectral index of -2.5 as suggested in the same reference, the MeV energy flux should be lower than 1$\times10^{-15}$~erg~s$^{-1}$~cm$^{-2}$, much lower than the flux we considered in our model and the sensitivity of next-generation MeV instruments.}  The distance to W44 is considered to be about 3 kpc according to the HI observations \citep{distance} and the detection of OH 1720 MHz maser spots implies the interaction between the SNR and the adjacent giant molecular cloud \citep{2005ApJ...627..803H}.  High-energy \grays possibly associated with W44 were first detected by EGRET \citep{EGRET_snr}. Then years later, AGILE and Fermi collaboration reported the characteristic pion-decay signature from W44 successively \citep{Agile2011W44,fermi_13}. Recently, a more detailed study performed by \citet{Peron2020}, which also preferred the hadronic origin of the \grays associated with W44, still cannot exclude the possible interpretation of radiation by electron bremsstrahlung due to the systematic uncertainties. In this work, we take W44 as a typical case to study the ability of future MeV telescopes to distinguish the pion-decay emission from the bremsstrahlung emission.

\begin{figure}
    \centering
    \begin{subfigure}[t]{1.0\linewidth}
        \centering
        \includegraphics[width=\linewidth]{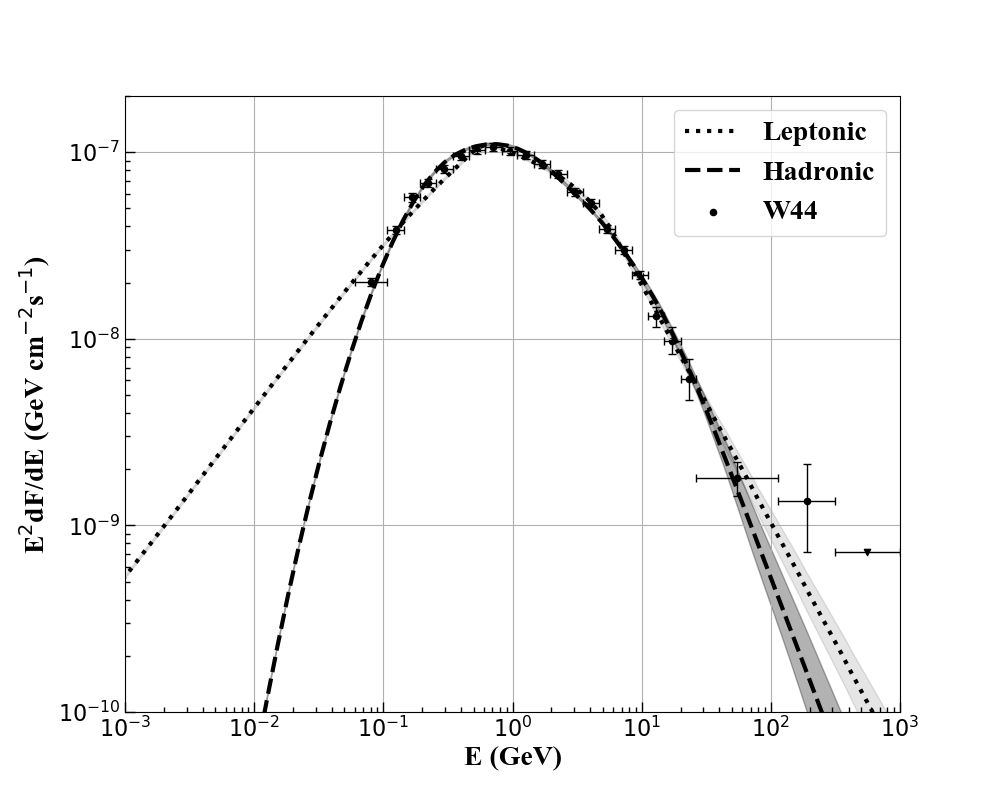}
        \label{fermi:1}
    \end{subfigure}
    \caption{SED of \grays from W44. The lines and shaded areas show the modeled flux and 1 $\sigma$ uncertainty of each scenario. The data points show the observed SED derived from Fermi Pass8 data\citep{Peron2020}.} \label{fig:fermi}
\end{figure}

Firstly,  we re-fitted the \gray SED (as shown in Fig. \ref{fig:fermi}) from the SNR W44 derived in \cite{Peron2020} with {\it naima} package \citep{naima}. { This Python package offers functionalities for calculating gamma-ray cross-sections and likelihood fitting of CR spectrum based on the observed gamma-ray spectrum.}
Here, we assumed two different origins of the \gray emission, interactions of accelerated protons (hadronic scenario) and nuclei with the surrounding gas, and bremsstrahlung from electrons (leptonic scenario). { There should also be IC emission from the same population of electrons that produce the radio/x-ray emission through synchrotron. However, in this case, the electrons producing 10-100~MeV gamma rays through IC scattering of optical/UV photons (dominate in this energy range), should have a Lorentz factor $\Gamma^2\sim10^7-10^8$. The corresponding synchrotron frequency of these electrons can be estimated as 1.3 $\Gamma^2(B/1\mathrm{\mu G})$ Hz. Assuming  $B\sim3~\mathrm{\mu G}$, the synchrotron frequency will be 0.1 -1 GHz. We also note that the energy density of the $3~\mathrm{\mu G}$ magnetic field is about 0.4~eV/cm$^3$, which is similar to the energy density of optical/UV photon fields in ISRF. Thus the energy flux of IC in 10-100~MeV should be similar to the synchrotron energy flux at 0.1 -1 GHz, which is 10$^{-15}$ erg/s/cm$^2$, again much smaller than the flux we considered in our model and the sensitivity of next-generation MeV instruments. Of course, the magnetic field can be much larger, but in this case, the IC flux should be even smaller. We conclude that the IC contribution is not significant and will not consider it in the following parts of the article. Furthermore, the current discrimination is based on the broken power-law model of both the electron and proton spectra. This model is also justified by ionization cooling of low-energy protons/electrons or by injection. However, it remains unclear whether there exists additional spectral structures at even lower energies, as no detections have been made in this energy range thus far. Therefore, we currently maintain the assumption of a broken power-law model for both the electron and proton spectra in our calculations.} We set the distribution of the parent protons or electrons as a broken power-law of momentum:
\begin{equation}
f(p)= \begin{cases}A\left(p / p_0\right)^{-\alpha_1} & : p<p_{\text {b }} \\ A\left(p_{\text {b }} / p_0\right)^{\alpha_2-\alpha_1}\left(p / p_0\right)^{-\alpha_2} & : p>p_{\text {b }}\end{cases}
\label{equ1}
\end{equation}
In Equ.\ref{equ1}, $p_\mathrm{b}$ is the break momentum, $A$ is the model amplitude at the break momentum, $p_0=$10 GeV/c,  $\alpha_1$ and $\alpha_2$ is the power law index for $p<p_\mathrm{b}$ and $p>p_\mathrm{b}$, respectively. A low energy cutoff at $p_{\min }^e=600~\mathrm{MeV}$/c was also added to the distribution function of the electrons in the leptonic scenario. 
The dashed line shows the fitted \gray emission in the hadronic scenario, and the dotted line shows the \gray spectrum obtained in the leptonic scenario. Meanwhile, the shaded area indicates the range of their 1$\sigma$ uncertainty. The fitted parameters for the distribution of protons and electrons in the two scenarios are shown in Tab.\ref{tab:para}. We found that with these parameters both leptonic and hadronic scenarios can explain the observed GeV SEDs, but in the MeV band, the flux can be different by orders of magnitude due to sharp spectral cutoff in the pion-decay gamma-ray spectrum.  Such differences can provide conclusive discrimination on these two scenarios with the observations in the MeV band.

\begin{table}
\caption{Parameters of the particle distribution in different scenarios.}  
\label{tab:para}       
\begin{tabular}{cccc}       
\hline\hline     
Scenario&$\alpha_1$&$\alpha_2$&$p_\mathrm{b}$ (GeV/c)\\
\hline
hadronic&2.44$\pm$0.03&3.78$\pm$0.15&35.5$\pm$4.1\\
leptonic&2.28$\pm$0.05&3.37$\pm$0.08&6.1$\pm$0.8\\
\hline
\end{tabular}
\end{table}

\section{Contribution from primary and secondary electrons in hadronic scenario}
\label{sec:pre}

In a hadronic scenario, the process $p p \rightarrow \pi^0 \rightarrow 2 \gamma$ dominates the \gray production. But even in the hadronic scenario, there will be inevitable primary electron accelerated at the same site of the CR protons, as well as secondary electrons produced accompanying the pion-decay \grays.  These primary and secondary electrons can also contribute to the overall \gray radiation, which can be even more significant below
100 MeV. Following the method of \cite{Yang2018}, we estimated the contribution from primary and secondary electrons to the \gray emission.

We first discuss the evolution of relativistic particles. In a given volume, this is given by the kinetic equation (see e.g., \cite{Ginzburg})
\begin{equation}
\frac{\partial N}{\partial t}=\frac{\partial}{\partial E}(P N)-\frac{N}{\tau_{\mathrm{esc}}}+Q
\end{equation}
where $P=P(E)=-\frac{d E}{d t}$ is the energy loss rate and $\tau_\mathrm{esc}$ is the characteristic escape time. Neglecting the particle escape from the \gray production region, we can give upper limits on the contribution of secondary electrons to the overall \gray flux.

For continuous injection $Q(E, t)=Q(E)$, the solution of kinetic equation becomes
\begin{equation}
\label{eq2}
N(E, t)=\frac{1}{P(E)} \int_E^{E_0} Q(E) \mathrm{d} E
\end{equation}
where $E_0$ is found by solving $t=\int_E^{E_0} \frac{\mathrm{d} E}{P(E)}$, which is the characteristic equation for the given epoch, $t$.

The cooling of electrons is mainly caused by ionization losses, synchrotron, bremsstrahlung, and IC radiation. The energy loss rate of ionisation $P_\mathrm{ion}$ is proportional to gas density, and scales as $1/\beta$ at low energies between 1 MeV and 1 GeV, where $\beta=v/c$. A convenient analytical presentation for the ionization losses can be found in \cite{Gould1972}. In the case of W44, we set the gas density $n=10~\mathrm{cm}^{-3}$\citep{Peron2020}, the magnetic fields $B=3~\mu$G, and the age $T=10^{12}~\mathrm{s}$.  The spectra of protons and electrons in different scenarios in Sec.2 are the evolved spectra derived from the observed gamma-ray data. As the break energy is as high as 35 GeV, the spectrum of protons in the MeV range is assumed to be power laws in momentum with the same index $\alpha_1$:
\begin{equation}
    Q(E)=\frac{N(p_0)}{\beta c}\left(\frac{p}{p_0}\right)^{-\alpha_1}
\end{equation}  
where $p$ is the proton momentum, $p_0$ is the reference point taken to be 10 GeV/c. 

\begin{figure}
    \centering
    \begin{subfigure}[t]{1.0\linewidth}
        \centering
        \includegraphics[width=\linewidth]{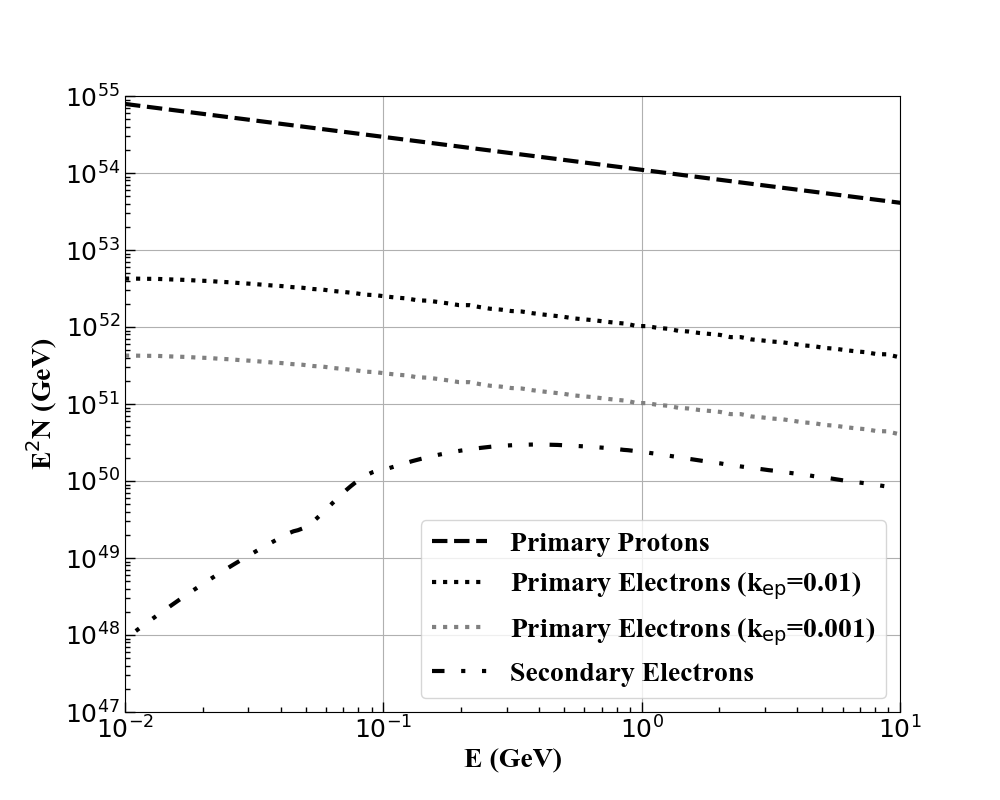}
        \label{N:1}
    \end{subfigure}
    \caption{Estimated energy distributions of the number of different particles after time revolution in the hadronic scenario. The gas density was set to be $n=$10 cm$^{-3}$ and the age was set to be $T=10^{12}$ s.} \label{fig:N}
\end{figure}

The interstellar radiation fields are assumed to comprise three components: the 2.7 K cosmic microwave background, whose energy density is 0.24 eV cm$^{-3}$,  the optical/UV field modeled as a grey-body component with an energy density of 2 eV cm$^{-3}$ and temperature of 5000 K,  and the infrared component which is modeled as a grey-body component with an energy density of 1 eV cm$^{-3}$ and a temperature of 100 K{\citep{Yang2018}}. For the primary electrons, it was assumed that their injection spectrum has the same shape as that of the primary protons. We regulated the injection spectrum of primary electrons by applying a constant e/p ratio ($k_{\rm ep}$) at 10 GeV. By calculating Eq.\ref{eq2}, we derived the spectrum of primary electrons after time evolution, which is shown in Fig.\ref{fig:N}. As for the secondary electrons, we assumed a constant distribution of parent protons to derive their injection spectrum. Then we calculated the spectrum after the time revolution by calculating Eq.\ref{eq2}, which is shown in Fig.\ref{fig:N}. We found that the contribution of secondary electrons can be ignored when compared to the primary electrons. Thus, only the contribution of primary electrons was considered for further calculation in Sec.4. However, this only applies to the case of W44. For other sources, due to differences in gas density, evolution time, and other factors, secondary electrons may still make a significant contribution \citep{Yang2018}.

\section{Detectability of next-generation MeV instruments}
\label{sec:detect}
Based on the calculations in Sec.2, the gamma-ray spectra produced in the hadronic and leptonic scenarios can exhibit significant differences in the MeV energy range. Therefore, we have estimated the future observational results of a next-generation MeV gamma-ray telescope, MeGaT, to evaluate its observational performance and determine whether its observations can distinguish between the radiation mechanisms that produce the \gray from W44. { As the simulation of the instruments is still in process, the instrument response function (IRF) and sensitivity curve of MeGaT will be presented in a dedicated work. In this article, we assumed that} MeGaT has an effective area of 100 cm$^2$, and its angular resolution was taken to be 1$^\circ$, { which is consistent with current predictions. Since the full IRF with energy dependence is not yet available, in the current work, as a schematic study,  we didn't take into account of the energy dependence of IRFs. }
The details of the use of these parameters during calculation are described later in this section.

\begin{figure}[h]
    \centering
    \includegraphics[width=1\linewidth]{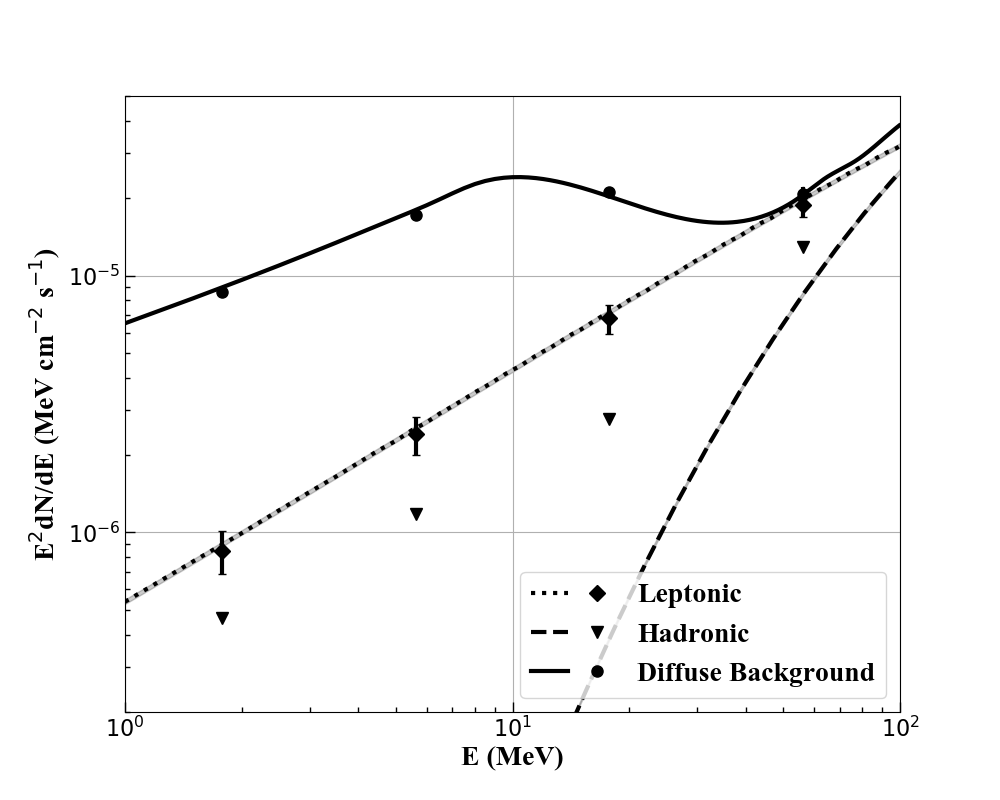}
    \caption{Estimated SED of future MeV \gray observations of W44 in different scenarios. This observation result adopted the instrument parameters of MeGaT, and $T_\mathrm{obs}$ was taken to be 2 months. The lines and shaded areas show the theoretical flux and 1 $\sigma$ uncertainty of each scenario. The data points and error bars show the derived flux and 1 $\sigma$ uncertainty of the four energy bins. The inverted triangles show the 3 $\sigma$ upper limit of the flux of \gray emission in the pion decay scenario.} \label{fig:2mon}
\end{figure}

Utilizing the fitted particle distributions obtained in Sec.2, we can calculate the \gray flux distributions with {\it naima} package in the MeV energy range corresponding to the two scenarios, respectively. { Although W44 is an extended source, its size is still much smaller than the PSFs of the MeV instruments. So we applied the "Aperture photometry" method and used all the photon counts in a disk region centered at W44 with a radius of the PSF(1$^\circ$) of MeGaT.} We also estimated the diffuse background flux in the direction of W44 by integrating the known MeV energy range data { in this disk region.}
For energy between 0.5 and 8.0 MeV, we adopted the diffuse Galactic emission re-analyzed by \cite{Siegert_2022} based on the observation of SPI aboard {\it INTEGRAL} using {\it GALPROP} \citep{galprop} models in a 95$^\circ~\times$ 95$^\circ$ region around the Galactic center. For energy over 50 MeV, we utilized the Fermi-LAT interstellar emission model \citep{fermibkg}, which was based on the first 9 years of Fermi-LAT data. A fits version of this model, {\it gll\_iem\_v07.fits}, is accessible on the Fermi Science Support Center website\footnote{\url{https://fermi.gsfc.nasa.gov/ssc/data/access/lat/BackgroundModels.html}}. Specially, we calculated the contribution of point sources in the same region by 14 years of the Fermi-LAT data using the {\it Fermitools} from Conda distribution\footnote{\url{https://github.com/fermi-lat/Fermitools-conda/}}. We generated the \gray counts maps and the exposure maps in different energy ranges to calculate the total flux in the reference region, and found that the results were similar to that from the Fermi interstellar emission model. However, in a larger integration area, 10$^\circ$ for example, the contribution of point sources would be nonnegligible due to the crowded source distribution on the Galactic plane.  The total flux would be two to three times higher than that of the diffuse background. Thus the angular resolution can be very important for such kind of study. The distribution of background flux in the whole energy range was then determined by cubic extrapolation in logarithmic space. Given the effective area ($A_\mathrm{eff}$) of different instruments, we calculated the total counts of \gray from W44 and the diffuse background in the 1-100 MeV energy range that could be observed within two months of observational time ($T_\mathrm{obs}$). We divided the data from 1 to 100 MeV into 4 energy bins. We estimated the possible  data counts of each bin assuming an exposure of two months by :
\begin{equation}
    N_\mathrm{counts}=\int^{E_\mathrm{upper}}_{E_\mathrm{lower}}F(E)T_\mathrm{obs}A_\mathrm{eff}(E)dE
\end{equation}
where $F(E)$ is the theoretical differential flux calculated in Sec.2, $T_\mathrm{obs}$ is the observational time, $A_\mathrm{eff}(E)$ is the effective area of the telescope, $E_\mathrm{lower}$ and $E_\mathrm{upper}$ are the lower and upper bond of the energy bin, respectively. In most of the energy range, the theoretical prediction of the \gray flux in the hadronic scenario is less than that of the leptonic scenario by more than one order of magnitude. Therefore, we only estimated the predicted counts of \gray in the leptonic scenario here. The results of the \gray counts are shown in Tab.\ref{tab:MeGaT}. We then assumed a Poisson distribution of the data and calculated the uncertainty $\sigma$ of the data by: 
\begin{equation}
    \sigma=\sqrt{N_\mathrm{total}}=\sqrt{N_\mathrm{signal}+N_\mathrm{bkg}}
\end{equation}
where $N_\mathrm{total}$ is the sum of the predicted \gray counts of signal ($N_\mathrm{signal}$) and background ($N_\mathrm{bkg}$). We also calculated the significance ($S$) of the data by:
\begin{equation}
    S=\frac{N_\mathrm{signal}}{\sqrt{N_\mathrm{signal}+N_\mathrm{bkg}}}
\end{equation}
As the counts of both signal and background are proportional to $T_\mathrm{obs}$, it naturally follows that the significance $S$ should be proportional to $\sqrt{T_\mathrm{obs}}$. The results are shown in Tab.\ref{tab:MeGaT} and Fig.\ref{fig:2mon}, which indicate that a $T_\mathrm{obs}$ of 2 months would be adequate for the data to reach enough significance.

To distinguish whether the \gray emission originated in a leptonic or hadronic scenario, we calculated the 3 $\sigma$ upper limit assuming the gamma rays were produced via the pion decay mechanism:
\begin{equation}
    F_\mathrm{up}=\frac{N_\mathrm{signal}+3\sqrt{N_\mathrm{signal}+N_\mathrm{bkg}}}{A_\mathrm{eff}(E_\mathrm{mid})T_\mathrm{obs}(E_\mathrm{upper}-E_\mathrm{lower})}
\end{equation}
where $F_\mathrm{up}$ is the 3 $\sigma$ upper limit of the measured \gray flux, $E_\mathrm{mid}=\sqrt{E_\mathrm{upper}E_\mathrm{lower}}$ is the central energy of the energy bin. The results are shown in Fig.\ref{fig:2mon}. As shown in the figure, we can easily distinguish between the two scenarios after two months of observation.

\begin{table}
\caption{Predicted counts of future MeGaT observation for \gray in the leptonic scenario ($T_\mathrm{obs}$ = 2 month).}  
\label{tab:MeGaT}       
\centering                        
\begin{tabular}{cccc}       
\hline\hline     
log(E/MeV)& $N_\mathrm{signal}$ &$N_\mathrm{bkg}$&S\\
&&&($\sigma$)\\
\hline
0.0-0.5&306&3099&5.24\\
0.5-1.0&274&1954&5.80\\
1.0-1.5&245&763&7.72\\
1.5-2.0&213&236&10.05\\
\hline
\end{tabular}
\end{table}

\begin{figure}
    \centering
    \begin{subfigure}[t]{1\linewidth}
        \centering
        \includegraphics[width=\linewidth]{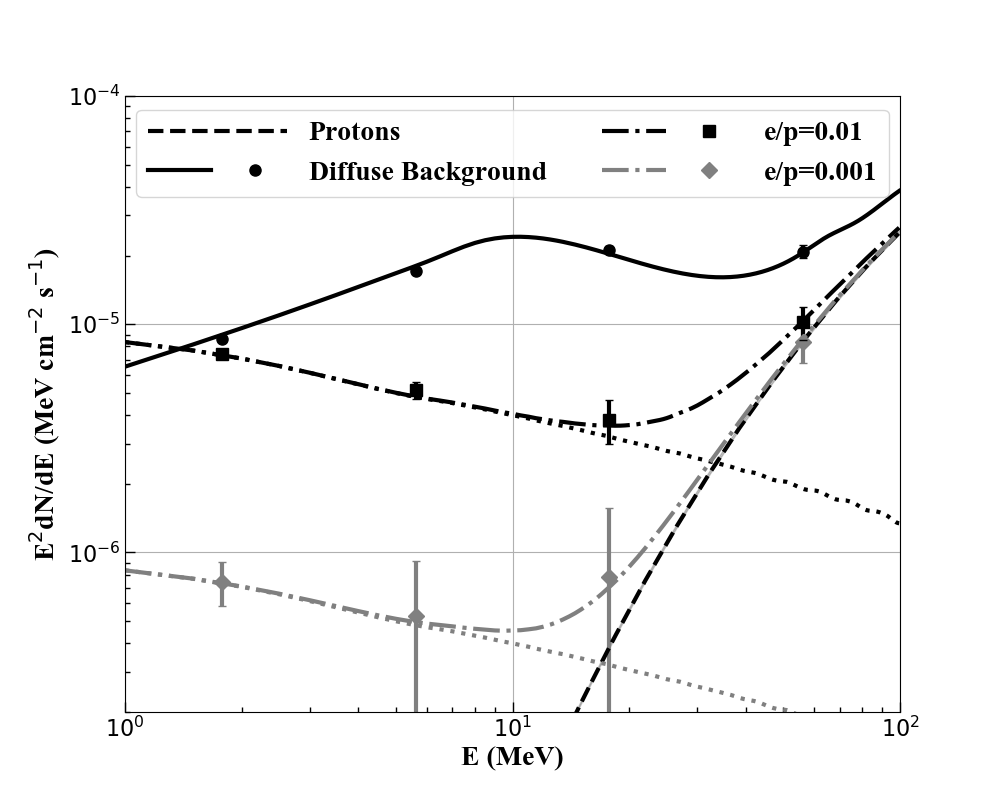}
        \label{2monpre:1}
    \end{subfigure}
    \caption{Estimated SED of future MeV \gray observations of W44 in { hadronic} scenario. This observation result adopted the instrument parameters of MeGaT, and $T_\mathrm{obs}$ was taken to be 2 months. The lines show the theoretical \gray flux from different origins. Specially, the black and grey dotted line shows the flux from the bremsstrahlung of primary electrons and the chain lines show the summed flux from both primary electrons and primary protons. The data points and error bars show the derived flux and 1 $\sigma$ uncertainty of the four energy bins.} \label{fig:pre}
\end{figure}

Besides, the MeV \gray observation can put limits on the e/p ratio in the hadronic scenario. Using the spectrum of primary electrons derived in Sec.3, we calculated the \gray contribution from the bremsstrahlung of these electrons and used the same method to calculate the telescope observation results in four energy bins. We assumed two uniform e/p ratios, 0.01 and 0.001, and applied similar calculations respectively. The results are shown in 
Tab.\ref{tab:pre} and Fig.\ref{fig:pre}. It can be seen that the change in the e/p ratio can significantly influence the MeV \gray spectrum, especially in lower energy. Due to the unavoidable contribution from primary electrons in the hadronic scenario, it would become more difficult to distinguish it from the leptonic model. But compared with the pure leptonic scenario, the hadronic scenario with primary electrons would introduce a distinct "valley" structure in the spectrum just below the pion bump (at dozens of MeV), as shown in Fig.\ref{fig:pre}. Such feature can also be recognized with an exposure of several months.

\begin{table}
\caption{Predicted counts of future MeGaT observation for \gray in the hadronic scenario ($T_\mathrm{obs}$ = 2 month). The counts of the diffuse background is the same with Tab.\ref{tab:MeGaT}.}  
\label{tab:pre}       
\centering                        
\begin{tabular}{ccccc}       
\hline\hline     
log(E/MeV)& \multicolumn{2}{c}{$N_\mathrm{signal}$} &\multicolumn{2}{c}{S ($\sigma$)}\\
&e/p=0.01&e/p=0.001&e/p=0.01&e/p=0.001\\
\hline
0.0-0.5&2676&268&35.21&4.62\\
0.5-1.0&583&60&11.57&1.34\\
1.0-1.5&137&28&4.57&1.00\\
1.5-2.0&116&95&6.19&5.23\\
\hline
\end{tabular}
\end{table}

\section{Discussion}
The pion-decay bump is decisive proof for the acceleration of CR protons in astrophysical objects. Such feature is most significant below 100 MeV and thus is an ideal scientific objective for next-generation MeV detectors. In this paper, we chose SNR W44 as an example to investigate the possible MeV \gray emission around and below the pion-decay bump. We found that, although the current \gray observations can be explained by both pion-decay and Bremsstrahlung processes, the future observations at about 10~MeV can distinguish these two mechanisms significantly with a reasonable exposure. 

{ We also explored alternative future MeV detectors such as The Compton Spectrometer and Imager(COSI)\citep{cosi}. Given its angular resolution of $4.1^{\circ}$ at $0.511 \mathrm{MeV}$ and $2.1^{\circ}$ at $1.809 \mathrm{MeV}$, we found that no significant number of other point sources within the region would fall within the angular resolution. The sensitivities shown in Fig.2 of \cite{cosi} also satisfy the requirements of the case of W44. With sufficient observational time, it should also possess the capability to identify the features outlined in our study. }

Furthermore, even in the pure hadronic case where the GeV \grays are all produced by the pion-decay process. There are inevitable primary electrons accelerated at the same site where CR protons are accelerated and also the secondary electrons from the decay of charged pions produced by the inelastic scattering of CR protons with the ambient gas. These primary and secondary electrons will also produce \grays via Bremsstrahlung. The \grays from these electrons are negligible at GeV energy range but can dominate below the pion-decay bump. In the case of W44, due to the relatively young age of this system, the contribution from secondary electrons is expected to be much smaller than those from primary electrons. Thus the precise measurement of the spectrum below the pion-decay bump in this source would provide unique information on the injected primary electron spectrum, especially can provide a direct measurement of the e/p ratio in this accelerator, which may provide important information to understand the particle acceleration as well as the origin of the anomalous electron spectrum observed locally \citep{pamela_e, atic_e,ams02_e,dampe_e}. 

 On the other hand, as calculated in \citet{Yang2018}, in the case when $n\times T$ is as large as $10^{15} ~\rm cm^{-3} s$ the bremsstrahlung \grays from secondary electrons can also contribute significantly, where $n$ is the ambient gas density and $T$ is the confinement time of the CR particles. Such conditions can be realized in dense regions near older accelerators. The measurement of MeV \grays in these environments can then provide  information on the particle confinement near the accelerators, which are believed to be more effective than in the ISM \citep{malkov12}. 

 To conclude, the MeV observation is a unique tool for probing CR-related science. For the specific case of SNR W44, in which the pion-decay bump feature was detected for the first time, the dedicated MeV investigations using future MeV detectors can significantly improve the ability to distinguish the pion-decay bump from the Bremsstrahlung emissions of CR electrons. Furthermore, in case the pion-decay bump is confirmed, the precise observation of \gray spectrum below the pion-decay bump can provide a direct measurement of e/p ratio in the accelerated CR in W44. Even after taking into account the diffuse background in this energy range, a marginal exposure of 2 months would be enough for such kind of study for planned MeV detectors such as MeGat and Amego \citep{amego2019}. For other older accelerators with higher ambient density, the MeV observations can also provide clues on the particle confinement near the accelerators. 
 
MeV \gray astronomy is an important window for CR study. For example, the direct measurement of MeV de-excitation nuclear lines induced by the inelastic scattering of low energy CRs (LECRs) with the ambient gas is probably the best way to study LECRs \citep{liu20,liu23}. The most accepted site for ultra-high energy CRs (VHECR/UHECRs) production is related to the supermassive black holes and associated powerful jet \citep{Hillas1984,2011ARA&A..49..119K}, in this dense environment the produced UHECRs will interact with the ambient medium and produce electromagnetic radiations. Due to the high opacity and effective pair production, only MeV \grays can escape and potentially be detected. Thus MeV observations are also crucial in understanding the origin of UHECRs\citep{2006NJPh....8..122D}. In this paper, in addition to the above two aspects, we showed that the MeV observations can also provide decisive criteria on the pion-decay nature of \gray emissions in the potential accelerator, and can be also used to study the e/p ratio in the accelerator as well as the confinement of CRs near the accelerators. These measurements can thus provide fresh information to understand the acceleration of CRs and could be another important scientific objective for future MeV astronomy.    

\section{Acknowledgements}

Rui-zhi Yang is supported by the NSFC under grant 12041305 and the national youth thousand talents program in China.
Bing Liu acknowledges the support from the NSFC under grant 12103049.

\bibliography{main.bib}

\begin{thebibliography}{33}%
\makeatletter
\providecommand \@ifxundefined [1]{%
 \@ifx{#1\undefined}
}%
\providecommand \@ifnum [1]{%
 \ifnum #1\expandafter \@firstoftwo
 \else \expandafter \@secondoftwo
 \fi
}%
\providecommand \@ifx [1]{%
 \ifx #1\expandafter \@firstoftwo
 \else \expandafter \@secondoftwo
 \fi
}%
\providecommand \natexlab [1]{#1}%
\providecommand \enquote  [1]{``#1''}%
\providecommand \bibnamefont  [1]{#1}%
\providecommand \bibfnamefont [1]{#1}%
\providecommand \citenamefont [1]{#1}%
\providecommand \href@noop [0]{\@secondoftwo}%
\providecommand \href [0]{\begingroup \@sanitize@url \@href}%
\providecommand \@href[1]{\@@startlink{#1}\@@href}%
\providecommand \@@href[1]{\endgroup#1\@@endlink}%
\providecommand \@sanitize@url [0]{\catcode `\\12\catcode `\$12\catcode
  `\&12\catcode `\#12\catcode `\^12\catcode `\_12\catcode `\%12\relax}%
\providecommand \@@startlink[1]{}%
\providecommand \@@endlink[0]{}%
\providecommand \url  [0]{\begingroup\@sanitize@url \@url }%
\providecommand \@url [1]{\endgroup\@href {#1}{\urlprefix }}%
\providecommand \urlprefix  [0]{URL }%
\providecommand \Eprint [0]{\href }%
\providecommand \doibase [0]{https://doi.org/}%
\providecommand \selectlanguage [0]{\@gobble}%
\providecommand \bibinfo  [0]{\@secondoftwo}%
\providecommand \bibfield  [0]{\@secondoftwo}%
\providecommand \translation [1]{[#1]}%
\providecommand \BibitemOpen [0]{}%
\providecommand \bibitemStop [0]{}%
\providecommand \bibitemNoStop [0]{.\EOS\space}%
\providecommand \EOS [0]{\spacefactor3000\relax}%
\providecommand \BibitemShut  [1]{\csname bibitem#1\endcsname}%
\let\auto@bib@innerbib\@empty
\bibitem [{\citenamefont {{Blasi}}(2013)}]{blasi13}%
  \BibitemOpen
  \bibfield  {author} {\bibinfo {author} {\bibfnamefont {P.}~\bibnamefont
  {{Blasi}}},\ }\bibfield  {title} {\bibinfo {title} {{The origin of galactic
  cosmic rays}},\ }\href {https://doi.org/10.1007/s00159-013-0070-7} {\bibfield
   {journal} {\bibinfo  {journal} {Astronomy and Astrophysics Review}\ }\textbf
  {\bibinfo {volume} {21}},\ \bibinfo {eid} {70} (\bibinfo {year} {2013})},\
  \Eprint {https://arxiv.org/abs/1311.7346} {arXiv:1311.7346 [astro-ph.HE]}
  \BibitemShut {NoStop}%
\bibitem [{\citenamefont {Acero}\ \emph {et~al.}(2016)\citenamefont {Acero},
  \citenamefont {Ackermann}, \citenamefont {Ajello} \emph
  {et~al.}}]{fermi_snr}%
  \BibitemOpen
  \bibfield  {author} {\bibinfo {author} {\bibfnamefont {F.}~\bibnamefont
  {Acero}}, \bibinfo {author} {\bibfnamefont {M.}~\bibnamefont {Ackermann}},
  \bibinfo {author} {\bibfnamefont {M.}~\bibnamefont {Ajello}}, \emph
  {et~al.},\ }\bibfield  {title} {\bibinfo {title} {The first fermi lat
  supernova remnant catalog},\ }\href
  {https://doi.org/10.3847/0067-0049/224/1/8} {\bibfield  {journal} {\bibinfo
  {journal} {The Astrophysical Journal Supplement Series}\ }\textbf {\bibinfo
  {volume} {224}},\ \bibinfo {pages} {8} (\bibinfo {year} {2016})}\BibitemShut
  {NoStop}%
\bibitem [{\citenamefont {{H.~E.~S.~S. Collaboration}}\ \emph
  {et~al.}(2018)\citenamefont {{H.~E.~S.~S. Collaboration}}, \citenamefont
  {{Abdalla}}, \citenamefont {{Abramowski}} \emph {et~al.}}]{hgps}%
  \BibitemOpen
  \bibfield  {author} {\bibinfo {author} {\bibnamefont {{H.~E.~S.~S.
  Collaboration}}}, \bibinfo {author} {\bibfnamefont {H.}~\bibnamefont
  {{Abdalla}}}, \bibinfo {author} {\bibfnamefont {A.}~\bibnamefont
  {{Abramowski}}}, \emph {et~al.},\ }\bibfield  {title} {\bibinfo {title} {{The
  H.E.S.S. Galactic plane survey}},\ }\href
  {https://doi.org/10.1051/0004-6361/201732098} {\bibfield  {journal} {\bibinfo
   {journal} {Astronomy and Astrophysics}\ }\textbf {\bibinfo {volume} {612}},\
  \bibinfo {eid} {A1} (\bibinfo {year} {2018})},\ \Eprint
  {https://arxiv.org/abs/1804.02432} {arXiv:1804.02432 [astro-ph.HE]}
  \BibitemShut {NoStop}%
\bibitem [{\citenamefont {{Giuliani}}\ \emph {et~al.}(2011)\citenamefont
  {{Giuliani}}, \citenamefont {{Cardillo}}, \citenamefont {{Tavani}} \emph
  {et~al.}}]{Agile2011W44}%
  \BibitemOpen
  \bibfield  {author} {\bibinfo {author} {\bibfnamefont {A.}~\bibnamefont
  {{Giuliani}}}, \bibinfo {author} {\bibfnamefont {M.}~\bibnamefont
  {{Cardillo}}}, \bibinfo {author} {\bibfnamefont {M.}~\bibnamefont
  {{Tavani}}}, \emph {et~al.},\ }\bibfield  {title} {\bibinfo {title} {{Neutral
  Pion Emission from Accelerated Protons in the Supernova Remnant W44}},\
  }\href {https://doi.org/10.1088/2041-8205/742/2/L30} {\bibfield  {journal}
  {\bibinfo  {journal} {The Astrophysical Journal}\ }\textbf {\bibinfo {volume}
  {742}},\ \bibinfo {eid} {L30} (\bibinfo {year} {2011})},\ \Eprint
  {https://arxiv.org/abs/1111.4868} {arXiv:1111.4868 [astro-ph.HE]}
  \BibitemShut {NoStop}%
\bibitem [{\citenamefont {{Ackermann}}\ \emph {et~al.}(2013)\citenamefont
  {{Ackermann}}, \citenamefont {{Ajello}}, \citenamefont {{Allafort}} \emph
  {et~al.}}]{fermi_13}%
  \BibitemOpen
  \bibfield  {author} {\bibinfo {author} {\bibfnamefont {M.}~\bibnamefont
  {{Ackermann}}}, \bibinfo {author} {\bibfnamefont {M.}~\bibnamefont
  {{Ajello}}}, \bibinfo {author} {\bibfnamefont {A.}~\bibnamefont
  {{Allafort}}}, \emph {et~al.},\ }\bibfield  {title} {\bibinfo {title}
  {{Detection of the Characteristic Pion-Decay Signature in Supernova
  Remnants}},\ }\href {https://doi.org/10.1126/science.1231160} {\bibfield
  {journal} {\bibinfo  {journal} {Science}\ }\textbf {\bibinfo {volume}
  {339}},\ \bibinfo {pages} {807} (\bibinfo {year} {2013})},\ \Eprint
  {https://arxiv.org/abs/1302.3307} {arXiv:1302.3307 [astro-ph.HE]}
  \BibitemShut {NoStop}%
\bibitem [{\citenamefont {{Peron}}\ \emph {et~al.}(2020)\citenamefont
  {{Peron}}, \citenamefont {{Aharonian}}, \citenamefont {{Casanova}} \emph
  {et~al.}}]{Peron2020}%
  \BibitemOpen
  \bibfield  {author} {\bibinfo {author} {\bibfnamefont {G.}~\bibnamefont
  {{Peron}}}, \bibinfo {author} {\bibfnamefont {F.}~\bibnamefont
  {{Aharonian}}}, \bibinfo {author} {\bibfnamefont {S.}~\bibnamefont
  {{Casanova}}}, \emph {et~al.},\ }\bibfield  {title} {\bibinfo {title} {{On
  the Gamma-Ray Emission of W44 and Its Surroundings}},\ }\href
  {https://doi.org/10.3847/2041-8213/ab93d1} {\bibfield  {journal} {\bibinfo
  {journal} {The Astrophysical Journal}\ }\textbf {\bibinfo {volume} {896}},\
  \bibinfo {eid} {L23} (\bibinfo {year} {2020})},\ \Eprint
  {https://arxiv.org/abs/2007.04821} {arXiv:2007.04821 [astro-ph.HE]}
  \BibitemShut {NoStop}%
\bibitem [{\citenamefont {{Strong}}\ \emph {et~al.}(2011)\citenamefont
  {{Strong}}, \citenamefont {{Orlando}},\ and\ \citenamefont
  {{Jaffe}}}]{strong11}%
  \BibitemOpen
  \bibfield  {author} {\bibinfo {author} {\bibfnamefont {A.~W.}\ \bibnamefont
  {{Strong}}}, \bibinfo {author} {\bibfnamefont {E.}~\bibnamefont
  {{Orlando}}},\ and\ \bibinfo {author} {\bibfnamefont {T.~R.}\ \bibnamefont
  {{Jaffe}}},\ }\bibfield  {title} {\bibinfo {title} {{The interstellar
  cosmic-ray electron spectrum from synchrotron radiation and direct
  measurements}},\ }\href {https://doi.org/10.1051/0004-6361/201116828}
  {\bibfield  {journal} {\bibinfo  {journal} {Astronomy and Astrophysics}\
  }\textbf {\bibinfo {volume} {534}},\ \bibinfo {eid} {A54} (\bibinfo {year}
  {2011})},\ \Eprint {https://arxiv.org/abs/1108.4822} {arXiv:1108.4822
  [astro-ph.HE]} \BibitemShut {NoStop}%
\bibitem [{\citenamefont {{Yang}}\ \emph {et~al.}(2018)\citenamefont {{Yang}},
  \citenamefont {{Kafexhiu}},\ and\ \citenamefont {{Aharonian}}}]{Yang2018}%
  \BibitemOpen
  \bibfield  {author} {\bibinfo {author} {\bibfnamefont {R.-z.}\ \bibnamefont
  {{Yang}}}, \bibinfo {author} {\bibfnamefont {E.}~\bibnamefont {{Kafexhiu}}},\
  and\ \bibinfo {author} {\bibfnamefont {F.}~\bibnamefont {{Aharonian}}},\
  }\bibfield  {title} {\bibinfo {title} {{Exploring the shape of the
  {\ensuremath{\gamma}}-ray spectrum around the
  ``{\ensuremath{\pi}}$^{0}$-bump''}},\ }\href
  {https://doi.org/10.1051/0004-6361/201730908} {\bibfield  {journal} {\bibinfo
   {journal} {Astronomy and Astrophysics}\ }\textbf {\bibinfo {volume} {615}},\
  \bibinfo {eid} {A108} (\bibinfo {year} {2018})},\ \Eprint
  {https://arxiv.org/abs/1803.05072} {arXiv:1803.05072 [astro-ph.HE]}
  \BibitemShut {NoStop}%
\bibitem [{\citenamefont {{Harrus}}\ \emph {et~al.}(1996)\citenamefont
  {{Harrus}}, \citenamefont {{Hughes}},\ and\ \citenamefont
  {{Helfand}}}]{1996ApJ...464L.161H}%
  \BibitemOpen
  \bibfield  {author} {\bibinfo {author} {\bibfnamefont {I.~M.}\ \bibnamefont
  {{Harrus}}}, \bibinfo {author} {\bibfnamefont {J.~P.}\ \bibnamefont
  {{Hughes}}},\ and\ \bibinfo {author} {\bibfnamefont {D.~J.}\ \bibnamefont
  {{Helfand}}},\ }\bibfield  {title} {\bibinfo {title} {{Discovery of an X-Ray
  Synchrotron Nebula Associated with the Radio Pulsar PSR B1853+01 in the
  Supernova Remnant W44}},\ }\href {https://doi.org/10.1086/310106} {\bibfield
  {journal} {\bibinfo  {journal} {The Astrophysical Journal}\ }\textbf
  {\bibinfo {volume} {464}},\ \bibinfo {pages} {L161} (\bibinfo {year}
  {1996})},\ \Eprint {https://arxiv.org/abs/astro-ph/9604120}
  {arXiv:astro-ph/9604120 [astro-ph]} \BibitemShut {NoStop}%
\bibitem [{\citenamefont {{Frail}}\ \emph {et~al.}(1996)\citenamefont
  {{Frail}}, \citenamefont {{Giacani}}, \citenamefont {{Goss}},\ and\
  \citenamefont {{Dubner}}}]{1996frail}%
  \BibitemOpen
  \bibfield  {author} {\bibinfo {author} {\bibfnamefont {D.~A.}\ \bibnamefont
  {{Frail}}}, \bibinfo {author} {\bibfnamefont {E.~B.}\ \bibnamefont
  {{Giacani}}}, \bibinfo {author} {\bibfnamefont {W.~M.}\ \bibnamefont
  {{Goss}}},\ and\ \bibinfo {author} {\bibfnamefont {G.}~\bibnamefont
  {{Dubner}}},\ }\bibfield  {title} {\bibinfo {title} {{The Pulsar Wind Nebula
  around PSR B1853+01 in the Supernova Remnant W44}},\ }\href
  {https://doi.org/10.1086/310103} {\bibfield  {journal} {\bibinfo  {journal}
  {The Astrophysical Journal Letters}\ }\textbf {\bibinfo {volume} {464}},\
  \bibinfo {pages} {L165} (\bibinfo {year} {1996})},\ \Eprint
  {https://arxiv.org/abs/astro-ph/9604121} {arXiv:astro-ph/9604121 [astro-ph]}
  \BibitemShut {NoStop}%
\bibitem [{\citenamefont {{Caswell}}\ \emph {et~al.}(1975)\citenamefont
  {{Caswell}}, \citenamefont {{Murray}}, \citenamefont {{Roger}} \emph
  {et~al.}}]{distance}%
  \BibitemOpen
  \bibfield  {author} {\bibinfo {author} {\bibfnamefont {J.~L.}\ \bibnamefont
  {{Caswell}}}, \bibinfo {author} {\bibfnamefont {J.~D.}\ \bibnamefont
  {{Murray}}}, \bibinfo {author} {\bibfnamefont {R.~S.}\ \bibnamefont
  {{Roger}}}, \emph {et~al.},\ }\bibfield  {title} {\bibinfo {title} {{Neutral
  hydrogen absorption measurements yielding kinematic distances for 42
  continuum sources in the galactic plane}},\ }\href@noop {} {\bibfield
  {journal} {\bibinfo  {journal} {Astronomy and Astrophysics}\ }\textbf
  {\bibinfo {volume} {45}},\ \bibinfo {pages} {239} (\bibinfo {year}
  {1975})}\BibitemShut {NoStop}%
\bibitem [{\citenamefont {{Hoffman}}\ \emph {et~al.}(2005)\citenamefont
  {{Hoffman}}, \citenamefont {{Goss}}, \citenamefont {{Brogan}},\ and\
  \citenamefont {{Claussen}}}]{2005ApJ...627..803H}%
  \BibitemOpen
  \bibfield  {author} {\bibinfo {author} {\bibfnamefont {I.~M.}\ \bibnamefont
  {{Hoffman}}}, \bibinfo {author} {\bibfnamefont {W.~M.}\ \bibnamefont
  {{Goss}}}, \bibinfo {author} {\bibfnamefont {C.~L.}\ \bibnamefont
  {{Brogan}}},\ and\ \bibinfo {author} {\bibfnamefont {M.~J.}\ \bibnamefont
  {{Claussen}}},\ }\bibfield  {title} {\bibinfo {title} {{The OH (1720 MHz)
  Supernova Remnant Masers in W44: MERLIN and VLBA Polarization
  Observations}},\ }\href {https://doi.org/10.1086/430419} {\bibfield
  {journal} {\bibinfo  {journal} {\apj}\ }\textbf {\bibinfo {volume} {627}},\
  \bibinfo {pages} {803} (\bibinfo {year} {2005})},\ \Eprint
  {https://arxiv.org/abs/astro-ph/0503481} {arXiv:astro-ph/0503481 [astro-ph]}
  \BibitemShut {NoStop}%
\bibitem [{\citenamefont {{Sturner}}\ and\ \citenamefont
  {{Dermer}}(1995)}]{EGRET_snr}%
  \BibitemOpen
  \bibfield  {author} {\bibinfo {author} {\bibfnamefont {S.~J.}\ \bibnamefont
  {{Sturner}}}\ and\ \bibinfo {author} {\bibfnamefont {C.~D.}\ \bibnamefont
  {{Dermer}}},\ }\bibfield  {title} {\bibinfo {title} {{Association of
  unidentified, low latitude EGRET sources with supernova remnants.}},\ }\href
  {https://doi.org/10.48550/arXiv.astro-ph/9409047} {\bibfield  {journal}
  {\bibinfo  {journal} {Astronomy and Astrophysics}\ }\textbf {\bibinfo
  {volume} {293}},\ \bibinfo {pages} {L17} (\bibinfo {year} {1995})},\ \Eprint
  {https://arxiv.org/abs/astro-ph/9409047} {arXiv:astro-ph/9409047 [astro-ph]}
  \BibitemShut {NoStop}%
\bibitem [{\citenamefont {{Zabalza}}(2015)}]{naima}%
  \BibitemOpen
  \bibfield  {author} {\bibinfo {author} {\bibfnamefont {V.}~\bibnamefont
  {{Zabalza}}},\ }\bibfield  {title} {\bibinfo {title} {naima: a python package
  for inference of relativistic particle energy distributions from observed
  nonthermal spectra},\ }\href@noop {} {\bibfield  {journal} {\bibinfo
  {journal} {Proc.~of International Cosmic Ray Conference 2015}\ ,\ \bibinfo
  {pages} {922}} (\bibinfo {year} {2015})},\ \Eprint
  {https://arxiv.org/abs/1509.03319} {1509.03319} \BibitemShut {NoStop}%
\bibitem [{\citenamefont {{Ginzburg}}\ and\ \citenamefont
  {{Syrovatskii}}(1964)}]{Ginzburg}%
  \BibitemOpen
  \bibfield  {author} {\bibinfo {author} {\bibfnamefont {V.~L.}\ \bibnamefont
  {{Ginzburg}}}\ and\ \bibinfo {author} {\bibfnamefont {S.~I.}\ \bibnamefont
  {{Syrovatskii}}},\ }\href@noop {} {\emph {\bibinfo {title} {{The Origin of
  Cosmic Rays}}}}\ (\bibinfo {year} {1964})\BibitemShut {NoStop}%
\bibitem [{\citenamefont {{Gould}}(1972)}]{Gould1972}%
  \BibitemOpen
  \bibfield  {author} {\bibinfo {author} {\bibfnamefont {R.~J.}\ \bibnamefont
  {{Gould}}},\ }\bibfield  {title} {\bibinfo {title} {{Energy loss of a
  relativistic ion in a plasma}},\ }\href
  {https://doi.org/10.1016/0031-8914(72)90159-0} {\bibfield  {journal}
  {\bibinfo  {journal} {Physica}\ }\textbf {\bibinfo {volume} {58}},\ \bibinfo
  {pages} {379} (\bibinfo {year} {1972})}\BibitemShut {NoStop}%
\bibitem [{\citenamefont {Siegert}\ \emph {et~al.}(2022)\citenamefont
  {Siegert}, \citenamefont {Berteaud}, \citenamefont {Calore}, \citenamefont
  {Serpico},\ and\ \citenamefont {Weinberger}}]{Siegert_2022}%
  \BibitemOpen
  \bibfield  {author} {\bibinfo {author} {\bibfnamefont {T.}~\bibnamefont
  {Siegert}}, \bibinfo {author} {\bibfnamefont {J.}~\bibnamefont {Berteaud}},
  \bibinfo {author} {\bibfnamefont {F.}~\bibnamefont {Calore}}, \bibinfo
  {author} {\bibfnamefont {P.~D.}\ \bibnamefont {Serpico}},\ and\ \bibinfo
  {author} {\bibfnamefont {C.}~\bibnamefont {Weinberger}},\ }\bibfield  {title}
  {\bibinfo {title} {Diffuse galactic emission spectrum between 0.5 and 8.0
  mev},\ }\href {https://doi.org/10.1051/0004-6361/202142639} {\bibfield
  {journal} {\bibinfo  {journal} {Astronomy \&amp; Astrophysics}\ }\textbf
  {\bibinfo {volume} {660}},\ \bibinfo {pages} {A130} (\bibinfo {year}
  {2022})}\BibitemShut {NoStop}%
\bibitem [{\citenamefont {{Vladimirov}}\ \emph {et~al.}(2011)\citenamefont
  {{Vladimirov}}, \citenamefont {{Digel}}, \citenamefont {{J{\'o}hannesson}}
  \emph {et~al.}}]{galprop}%
  \BibitemOpen
  \bibfield  {author} {\bibinfo {author} {\bibfnamefont {A.~E.}\ \bibnamefont
  {{Vladimirov}}}, \bibinfo {author} {\bibfnamefont {S.~W.}\ \bibnamefont
  {{Digel}}}, \bibinfo {author} {\bibfnamefont {G.}~\bibnamefont
  {{J{\'o}hannesson}}}, \emph {et~al.},\ }\bibfield  {title} {\bibinfo {title}
  {{GALPROP WebRun: An internet-based service for calculating galactic cosmic
  ray propagation and associated photon emissions}},\ }\href
  {https://doi.org/10.1016/j.cpc.2011.01.017} {\bibfield  {journal} {\bibinfo
  {journal} {Computer Physics Communications}\ }\textbf {\bibinfo {volume}
  {182}},\ \bibinfo {pages} {1156} (\bibinfo {year} {2011})},\ \Eprint
  {https://arxiv.org/abs/1008.3642} {arXiv:1008.3642 [astro-ph.HE]}
  \BibitemShut {NoStop}%
\bibitem [{\citenamefont {collaboration}(2019)}]{fermibkg}%
  \BibitemOpen
  \bibfield  {author} {\bibinfo {author} {\bibfnamefont {T.~F.-L.}\
  \bibnamefont {collaboration}},\ }\bibfield  {title} {\bibinfo {title}
  {Galactic interstellar emission model for the 4fgl catalog analysis},\ }\href
  {https://fermi.gsfc.nasa.gov/ssc/data/analysis/software/aux/4fgl/Galactic_Diffuse_Emission_Model_for_the_4FGL_Catalog_Analysis.pdf}
  {\bibfield  {journal} {\bibinfo  {journal} {open document}\ } (\bibinfo
  {year} {2019})}\BibitemShut {NoStop}%
\bibitem [{Note1()}]{Note1}%
  \BibitemOpen
  \bibinfo {note} {\protect \url
  {https://fermi.gsfc.nasa.gov/ssc/data/access/lat/BackgroundModels.html}}\BibitemShut
  {NoStop}%
\bibitem [{Note2()}]{Note2}%
  \BibitemOpen
  \bibinfo {note} {\protect \url
  {https://github.com/fermi-lat/Fermitools-conda/}}\BibitemShut {NoStop}%
\bibitem [{\citenamefont {{Tomsick}}\ \emph {et~al.}(2023)\citenamefont
  {{Tomsick}}, \citenamefont {{Boggs}}, \citenamefont {{Zoglauer}},
  \citenamefont {{Hartmann}}, \citenamefont {{Ajello}}, \citenamefont
  {{Burns}}, \citenamefont {{Fryer}}, \citenamefont {{Karwin}}, \citenamefont
  {{Kierans}}, \citenamefont {{Lowell}}, \citenamefont {{Malzac}},
  \citenamefont {{Roberts}}, \citenamefont {{Saint-Hilaire}}, \citenamefont
  {{Shih}}, \citenamefont {{Siegert}}, \citenamefont {{Sleator}}, \citenamefont
  {{Takahashi}}, \citenamefont {{Tavecchio}}, \citenamefont {{Wulf}},
  \citenamefont {{Beechert}}, \citenamefont {{Gulick}}, \citenamefont
  {{Joens}}, \citenamefont {{Lazar}}, \citenamefont {{Neights}}, \citenamefont
  {{Martinez Oliveros}}, \citenamefont {{Matsumoto}}, \citenamefont {{Melia}},
  \citenamefont {{Yoneda}}, \citenamefont {{Amman}}, \citenamefont {{Bal}},
  \citenamefont {{von Ballmoos}}, \citenamefont {{Bates}}, \citenamefont
  {{B{\"o}ttcher}}, \citenamefont {{Bulgarelli}}, \citenamefont {{Cavazzuti}},
  \citenamefont {{Chang}}, \citenamefont {{Chen}}, \citenamefont {{Chu}},
  \citenamefont {{Ciabattoni}}, \citenamefont {{Costamante}}, \citenamefont
  {{Dreyer}}, \citenamefont {{Fioretti}}, \citenamefont {{Fenu}}, \citenamefont
  {{Gallego}}, \citenamefont {{Ghirlanda}}, \citenamefont {{Grove}},
  \citenamefont {{Huang}}, \citenamefont {{Jean}}, \citenamefont {{Khatiya}},
  \citenamefont {{Kn{\"o}dlseder}}, \citenamefont {{Krause}}, \citenamefont
  {{Leising}}, \citenamefont {{Lewis}}, \citenamefont {{Lommler}},
  \citenamefont {{Marcotulli}}, \citenamefont {{Martinez-Castellanos}},
  \citenamefont {{Mittal}}, \citenamefont {{Negro}}, \citenamefont {{Al
  Nussirat}}, \citenamefont {{Nakazawa}}, \citenamefont {{Oberlack}},
  \citenamefont {{Palmore}}, \citenamefont {{Panebianco}}, \citenamefont
  {{Parmiggiani}}, \citenamefont {{Parsotan}}, \citenamefont {{Pike}},
  \citenamefont {{Rogers}}, \citenamefont {{Schutte}}, \citenamefont {{Sheng}},
  \citenamefont {{Smale}}, \citenamefont {{Smith}}, \citenamefont {{Trigg}},
  \citenamefont {{Venters}}, \citenamefont {{Watanabe}},\ and\ \citenamefont
  {{Zhang}}}]{cosi}%
  \BibitemOpen
  \bibfield  {author} {\bibinfo {author} {\bibfnamefont {J.~A.}\ \bibnamefont
  {{Tomsick}}}, \bibinfo {author} {\bibfnamefont {S.~E.}\ \bibnamefont
  {{Boggs}}}, \bibinfo {author} {\bibfnamefont {A.}~\bibnamefont {{Zoglauer}}},
  \bibinfo {author} {\bibfnamefont {D.}~\bibnamefont {{Hartmann}}}, \bibinfo
  {author} {\bibfnamefont {M.}~\bibnamefont {{Ajello}}}, \bibinfo {author}
  {\bibfnamefont {E.}~\bibnamefont {{Burns}}}, \bibinfo {author} {\bibfnamefont
  {C.}~\bibnamefont {{Fryer}}}, \bibinfo {author} {\bibfnamefont
  {C.}~\bibnamefont {{Karwin}}}, \bibinfo {author} {\bibfnamefont
  {C.}~\bibnamefont {{Kierans}}}, \bibinfo {author} {\bibfnamefont
  {A.}~\bibnamefont {{Lowell}}}, \bibinfo {author} {\bibfnamefont
  {J.}~\bibnamefont {{Malzac}}}, \bibinfo {author} {\bibfnamefont
  {J.}~\bibnamefont {{Roberts}}}, \bibinfo {author} {\bibfnamefont
  {P.}~\bibnamefont {{Saint-Hilaire}}}, \bibinfo {author} {\bibfnamefont
  {A.}~\bibnamefont {{Shih}}}, \bibinfo {author} {\bibfnamefont
  {T.}~\bibnamefont {{Siegert}}}, \bibinfo {author} {\bibfnamefont
  {C.}~\bibnamefont {{Sleator}}}, \bibinfo {author} {\bibfnamefont
  {T.}~\bibnamefont {{Takahashi}}}, \bibinfo {author} {\bibfnamefont
  {F.}~\bibnamefont {{Tavecchio}}}, \bibinfo {author} {\bibfnamefont
  {E.}~\bibnamefont {{Wulf}}}, \bibinfo {author} {\bibfnamefont
  {J.}~\bibnamefont {{Beechert}}}, \bibinfo {author} {\bibfnamefont
  {H.}~\bibnamefont {{Gulick}}}, \bibinfo {author} {\bibfnamefont
  {A.}~\bibnamefont {{Joens}}}, \bibinfo {author} {\bibfnamefont
  {H.}~\bibnamefont {{Lazar}}}, \bibinfo {author} {\bibfnamefont
  {E.}~\bibnamefont {{Neights}}}, \bibinfo {author} {\bibfnamefont {J.~C.}\
  \bibnamefont {{Martinez Oliveros}}}, \bibinfo {author} {\bibfnamefont
  {S.}~\bibnamefont {{Matsumoto}}}, \bibinfo {author} {\bibfnamefont
  {T.}~\bibnamefont {{Melia}}}, \bibinfo {author} {\bibfnamefont
  {H.}~\bibnamefont {{Yoneda}}}, \bibinfo {author} {\bibfnamefont
  {M.}~\bibnamefont {{Amman}}}, \bibinfo {author} {\bibfnamefont
  {D.}~\bibnamefont {{Bal}}}, \bibinfo {author} {\bibfnamefont
  {P.}~\bibnamefont {{von Ballmoos}}}, \bibinfo {author} {\bibfnamefont
  {H.}~\bibnamefont {{Bates}}}, \bibinfo {author} {\bibfnamefont
  {M.}~\bibnamefont {{B{\"o}ttcher}}}, \bibinfo {author} {\bibfnamefont
  {A.}~\bibnamefont {{Bulgarelli}}}, \bibinfo {author} {\bibfnamefont
  {E.}~\bibnamefont {{Cavazzuti}}}, \bibinfo {author} {\bibfnamefont {H.-K.}\
  \bibnamefont {{Chang}}}, \bibinfo {author} {\bibfnamefont {C.}~\bibnamefont
  {{Chen}}}, \bibinfo {author} {\bibfnamefont {C.-Y.}\ \bibnamefont {{Chu}}},
  \bibinfo {author} {\bibfnamefont {A.}~\bibnamefont {{Ciabattoni}}}, \bibinfo
  {author} {\bibfnamefont {L.}~\bibnamefont {{Costamante}}}, \bibinfo {author}
  {\bibfnamefont {L.}~\bibnamefont {{Dreyer}}}, \bibinfo {author}
  {\bibfnamefont {V.}~\bibnamefont {{Fioretti}}}, \bibinfo {author}
  {\bibfnamefont {F.}~\bibnamefont {{Fenu}}}, \bibinfo {author} {\bibfnamefont
  {S.}~\bibnamefont {{Gallego}}}, \bibinfo {author} {\bibfnamefont
  {G.}~\bibnamefont {{Ghirlanda}}}, \bibinfo {author} {\bibfnamefont
  {E.}~\bibnamefont {{Grove}}}, \bibinfo {author} {\bibfnamefont {C.-Y.}\
  \bibnamefont {{Huang}}}, \bibinfo {author} {\bibfnamefont {P.}~\bibnamefont
  {{Jean}}}, \bibinfo {author} {\bibfnamefont {N.}~\bibnamefont {{Khatiya}}},
  \bibinfo {author} {\bibfnamefont {J.}~\bibnamefont {{Kn{\"o}dlseder}}},
  \bibinfo {author} {\bibfnamefont {M.}~\bibnamefont {{Krause}}}, \bibinfo
  {author} {\bibfnamefont {M.}~\bibnamefont {{Leising}}}, \bibinfo {author}
  {\bibfnamefont {T.~R.}\ \bibnamefont {{Lewis}}}, \bibinfo {author}
  {\bibfnamefont {J.~P.}\ \bibnamefont {{Lommler}}}, \bibinfo {author}
  {\bibfnamefont {L.}~\bibnamefont {{Marcotulli}}}, \bibinfo {author}
  {\bibfnamefont {I.}~\bibnamefont {{Martinez-Castellanos}}}, \bibinfo {author}
  {\bibfnamefont {S.}~\bibnamefont {{Mittal}}}, \bibinfo {author}
  {\bibfnamefont {M.}~\bibnamefont {{Negro}}}, \bibinfo {author} {\bibfnamefont
  {S.}~\bibnamefont {{Al Nussirat}}}, \bibinfo {author} {\bibfnamefont
  {K.}~\bibnamefont {{Nakazawa}}}, \bibinfo {author} {\bibfnamefont
  {U.}~\bibnamefont {{Oberlack}}}, \bibinfo {author} {\bibfnamefont
  {D.}~\bibnamefont {{Palmore}}}, \bibinfo {author} {\bibfnamefont
  {G.}~\bibnamefont {{Panebianco}}}, \bibinfo {author} {\bibfnamefont
  {N.}~\bibnamefont {{Parmiggiani}}}, \bibinfo {author} {\bibfnamefont
  {T.}~\bibnamefont {{Parsotan}}}, \bibinfo {author} {\bibfnamefont {S.~N.}\
  \bibnamefont {{Pike}}}, \bibinfo {author} {\bibfnamefont {F.}~\bibnamefont
  {{Rogers}}}, \bibinfo {author} {\bibfnamefont {H.}~\bibnamefont {{Schutte}}},
  \bibinfo {author} {\bibfnamefont {Y.}~\bibnamefont {{Sheng}}}, \bibinfo
  {author} {\bibfnamefont {A.~P.}\ \bibnamefont {{Smale}}}, \bibinfo {author}
  {\bibfnamefont {J.}~\bibnamefont {{Smith}}}, \bibinfo {author} {\bibfnamefont
  {A.}~\bibnamefont {{Trigg}}}, \bibinfo {author} {\bibfnamefont
  {T.}~\bibnamefont {{Venters}}}, \bibinfo {author} {\bibfnamefont
  {Y.}~\bibnamefont {{Watanabe}}},\ and\ \bibinfo {author} {\bibfnamefont
  {H.}~\bibnamefont {{Zhang}}},\ }\bibfield  {title} {\bibinfo {title} {{The
  Compton Spectrometer and Imager}},\ }\href
  {https://doi.org/10.48550/arXiv.2308.12362} {\bibfield  {journal} {\bibinfo
  {journal} {arXiv e-prints}\ ,\ \bibinfo {eid} {arXiv:2308.12362}} (\bibinfo
  {year} {2023})},\ \Eprint {https://arxiv.org/abs/2308.12362}
  {arXiv:2308.12362 [astro-ph.HE]} \BibitemShut {NoStop}%
\bibitem [{\citenamefont {{Adriani}}\ \emph {et~al.}(2009)\citenamefont
  {{Adriani}}, \citenamefont {{Barbarino}}, \citenamefont {{Bazilevskaya}}
  \emph {et~al.}}]{pamela_e}%
  \BibitemOpen
  \bibfield  {author} {\bibinfo {author} {\bibfnamefont {O.}~\bibnamefont
  {{Adriani}}}, \bibinfo {author} {\bibfnamefont {G.~C.}\ \bibnamefont
  {{Barbarino}}}, \bibinfo {author} {\bibfnamefont {G.~A.}\ \bibnamefont
  {{Bazilevskaya}}}, \emph {et~al.},\ }\bibfield  {title} {\bibinfo {title}
  {{An anomalous positron abundance in cosmic rays with energies 1.5-100GeV}},\
  }\href {https://doi.org/10.1038/nature07942} {\bibfield  {journal} {\bibinfo
  {journal} {\nat}\ }\textbf {\bibinfo {volume} {458}},\ \bibinfo {pages} {607}
  (\bibinfo {year} {2009})},\ \Eprint {https://arxiv.org/abs/0810.4995}
  {arXiv:0810.4995 [astro-ph]} \BibitemShut {NoStop}%
\bibitem [{\citenamefont {{Chang}}\ \emph {et~al.}(2008)\citenamefont
  {{Chang}}, \citenamefont {{Adams}}, \citenamefont {{Ahn}} \emph
  {et~al.}}]{atic_e}%
  \BibitemOpen
  \bibfield  {author} {\bibinfo {author} {\bibfnamefont {J.}~\bibnamefont
  {{Chang}}}, \bibinfo {author} {\bibfnamefont {J.~H.}\ \bibnamefont
  {{Adams}}}, \bibinfo {author} {\bibfnamefont {H.~S.}\ \bibnamefont {{Ahn}}},
  \emph {et~al.},\ }\bibfield  {title} {\bibinfo {title} {{An excess of cosmic
  ray electrons at energies of 300-800GeV}},\ }\href
  {https://doi.org/10.1038/nature07477} {\bibfield  {journal} {\bibinfo
  {journal} {\nat}\ }\textbf {\bibinfo {volume} {456}},\ \bibinfo {pages} {362}
  (\bibinfo {year} {2008})}\BibitemShut {NoStop}%
\bibitem [{\citenamefont {{Aguilar}}\ \emph {et~al.}(2014)\citenamefont
  {{Aguilar}}, \citenamefont {{Aisa}}, \citenamefont {{Alvino}} \emph
  {et~al.}}]{ams02_e}%
  \BibitemOpen
  \bibfield  {author} {\bibinfo {author} {\bibfnamefont {M.}~\bibnamefont
  {{Aguilar}}}, \bibinfo {author} {\bibfnamefont {D.}~\bibnamefont {{Aisa}}},
  \bibinfo {author} {\bibfnamefont {A.}~\bibnamefont {{Alvino}}}, \emph
  {et~al.},\ }\bibfield  {title} {\bibinfo {title} {{Electron and Positron
  Fluxes in Primary Cosmic Rays Measured with the Alpha Magnetic Spectrometer
  on the International Space Station}},\ }\href
  {https://doi.org/10.1103/PhysRevLett.113.121102} {\bibfield  {journal}
  {\bibinfo  {journal} {\prl}\ }\textbf {\bibinfo {volume} {113}},\ \bibinfo
  {eid} {121102} (\bibinfo {year} {2014})}\BibitemShut {NoStop}%
\bibitem [{\citenamefont {{DAMPE Collaboration}}\ \emph
  {et~al.}(2017)\citenamefont {{DAMPE Collaboration}}, \citenamefont
  {{Ambrosi}}, \citenamefont {{An}} \emph {et~al.}}]{dampe_e}%
  \BibitemOpen
  \bibfield  {author} {\bibinfo {author} {\bibnamefont {{DAMPE
  Collaboration}}}, \bibinfo {author} {\bibfnamefont {G.}~\bibnamefont
  {{Ambrosi}}}, \bibinfo {author} {\bibfnamefont {Q.}~\bibnamefont {{An}}},
  \emph {et~al.},\ }\bibfield  {title} {\bibinfo {title} {{Direct detection of
  a break in the teraelectronvolt cosmic-ray spectrum of electrons and
  positrons}},\ }\href {https://doi.org/10.1038/nature24475} {\bibfield
  {journal} {\bibinfo  {journal} {\nat}\ }\textbf {\bibinfo {volume} {552}},\
  \bibinfo {pages} {63} (\bibinfo {year} {2017})},\ \Eprint
  {https://arxiv.org/abs/1711.10981} {arXiv:1711.10981 [astro-ph.HE]}
  \BibitemShut {NoStop}%
\bibitem [{\citenamefont {{Malkov}}\ \emph {et~al.}(2012)\citenamefont
  {{Malkov}}, \citenamefont {{Diamond}},\ and\ \citenamefont
  {{Sagdeev}}}]{malkov12}%
  \BibitemOpen
  \bibfield  {author} {\bibinfo {author} {\bibfnamefont {M.~A.}\ \bibnamefont
  {{Malkov}}}, \bibinfo {author} {\bibfnamefont {P.~H.}\ \bibnamefont
  {{Diamond}}},\ and\ \bibinfo {author} {\bibfnamefont {R.~Z.}\ \bibnamefont
  {{Sagdeev}}},\ }\bibfield  {title} {\bibinfo {title} {{On the mechanism for
  breaks in the cosmic ray spectruma)}},\ }\href
  {https://doi.org/10.1063/1.4737584} {\bibfield  {journal} {\bibinfo
  {journal} {Physics of Plasmas}\ }\textbf {\bibinfo {volume} {19}},\ \bibinfo
  {eid} {082901} (\bibinfo {year} {2012})},\ \Eprint
  {https://arxiv.org/abs/1206.1384} {arXiv:1206.1384 [astro-ph.GA]}
  \BibitemShut {NoStop}%
\bibitem [{\citenamefont {{McEnery}}\ \emph {et~al.}(2019)\citenamefont
  {{McEnery}}, \citenamefont {{van der Horst}}, \citenamefont {{Dominguez}}
  \emph {et~al.}}]{amego2019}%
  \BibitemOpen
  \bibfield  {author} {\bibinfo {author} {\bibfnamefont {J.}~\bibnamefont
  {{McEnery}}}, \bibinfo {author} {\bibfnamefont {A.}~\bibnamefont {{van der
  Horst}}}, \bibinfo {author} {\bibfnamefont {A.}~\bibnamefont {{Dominguez}}},
  \emph {et~al.},\ }\bibfield  {title} {\bibinfo {title} {{All-sky Medium
  Energy Gamma-ray Observatory: Exploring the Extreme Multimessenger
  Universe}},\ }in\ \href@noop {} {\emph {\bibinfo {booktitle} {Bulletin of the
  American Astronomical Society}}},\ Vol.~\bibinfo {volume} {51}\ (\bibinfo
  {year} {2019})\ p.\ \bibinfo {pages} {245},\ \Eprint
  {https://arxiv.org/abs/1907.07558} {arXiv:1907.07558 [astro-ph.IM]}
  \BibitemShut {NoStop}%
\bibitem [{\citenamefont {{Liu}}\ \emph {et~al.}(2021)\citenamefont {{Liu}},
  \citenamefont {{Yang}},\ and\ \citenamefont {{Aharonian}}}]{liu20}%
  \BibitemOpen
  \bibfield  {author} {\bibinfo {author} {\bibfnamefont {B.}~\bibnamefont
  {{Liu}}}, \bibinfo {author} {\bibfnamefont {R.-z.}\ \bibnamefont {{Yang}}},\
  and\ \bibinfo {author} {\bibfnamefont {F.}~\bibnamefont {{Aharonian}}},\
  }\bibfield  {title} {\bibinfo {title} {{Nuclear de-excitation lines as a
  probe of low-energy cosmic rays}},\ }\href
  {https://doi.org/10.1051/0004-6361/202039977} {\bibfield  {journal} {\bibinfo
   {journal} {Astronomy and Astrophysics}\ }\textbf {\bibinfo {volume} {646}},\
  \bibinfo {eid} {A149} (\bibinfo {year} {2021})},\ \Eprint
  {https://arxiv.org/abs/2101.03695} {arXiv:2101.03695 [astro-ph.HE]}
  \BibitemShut {NoStop}%
\bibitem [{\citenamefont {{Liu}}\ \emph {et~al.}(2023)\citenamefont {{Liu}},
  \citenamefont {{Yang}}, \citenamefont {{He}},\ and\ \citenamefont
  {{Aharonian}}}]{liu23}%
  \BibitemOpen
  \bibfield  {author} {\bibinfo {author} {\bibfnamefont {B.}~\bibnamefont
  {{Liu}}}, \bibinfo {author} {\bibfnamefont {R.-z.}\ \bibnamefont {{Yang}}},
  \bibinfo {author} {\bibfnamefont {X.-y.}\ \bibnamefont {{He}}},\ and\
  \bibinfo {author} {\bibfnamefont {F.}~\bibnamefont {{Aharonian}}},\
  }\bibfield  {title} {\bibinfo {title} {{New estimation of the nuclear
  de-excitation line emission from the supernova remnant Cassiopeia A}},\
  }\href {https://doi.org/10.1093/mnras/stad2165} {\bibfield  {journal}
  {\bibinfo  {journal} {Monthly Notices of the Royal Astronomical Society}\
  }\textbf {\bibinfo {volume} {524}},\ \bibinfo {pages} {5248} (\bibinfo {year}
  {2023})},\ \Eprint {https://arxiv.org/abs/2307.08967} {arXiv:2307.08967
  [astro-ph.HE]} \BibitemShut {NoStop}%
\bibitem [{\citenamefont {{Hillas}}(1984)}]{Hillas1984}%
  \BibitemOpen
  \bibfield  {author} {\bibinfo {author} {\bibfnamefont {A.~M.}\ \bibnamefont
  {{Hillas}}},\ }\bibfield  {title} {\bibinfo {title} {{The Origin of
  Ultra-High-Energy Cosmic Rays}},\ }\href
  {https://doi.org/10.1146/annurev.aa.22.090184.002233} {\bibfield  {journal}
  {\bibinfo  {journal} {Annual Review of Astronomy and Astrophysics}\ }\textbf
  {\bibinfo {volume} {22}},\ \bibinfo {pages} {425} (\bibinfo {year}
  {1984})}\BibitemShut {NoStop}%
\bibitem [{\citenamefont {{Kotera}}\ and\ \citenamefont
  {{Olinto}}(2011)}]{2011ARA&A..49..119K}%
  \BibitemOpen
  \bibfield  {author} {\bibinfo {author} {\bibfnamefont {K.}~\bibnamefont
  {{Kotera}}}\ and\ \bibinfo {author} {\bibfnamefont {A.~V.}\ \bibnamefont
  {{Olinto}}},\ }\bibfield  {title} {\bibinfo {title} {{The Astrophysics of
  Ultrahigh-Energy Cosmic Rays}},\ }\href
  {https://doi.org/10.1146/annurev-astro-081710-102620} {\bibfield  {journal}
  {\bibinfo  {journal} {Annual Review of Astronomy and Astrophysics}\ }\textbf
  {\bibinfo {volume} {49}},\ \bibinfo {pages} {119} (\bibinfo {year} {2011})},\
  \Eprint {https://arxiv.org/abs/1101.4256} {arXiv:1101.4256 [astro-ph.HE]}
  \BibitemShut {NoStop}%
\bibitem [{\citenamefont {{Dermer}}\ and\ \citenamefont
  {{Atoyan}}(2006)}]{2006NJPh....8..122D}%
  \BibitemOpen
  \bibfield  {author} {\bibinfo {author} {\bibfnamefont {C.~D.}\ \bibnamefont
  {{Dermer}}}\ and\ \bibinfo {author} {\bibfnamefont {A.}~\bibnamefont
  {{Atoyan}}},\ }\bibfield  {title} {\bibinfo {title} {{Ultra-high energy
  cosmic rays, cascade gamma rays, and high-energy neutrinos from gamma-ray
  bursts}},\ }\href {https://doi.org/10.1088/1367-2630/8/7/122} {\bibfield
  {journal} {\bibinfo  {journal} {New Journal of Physics}\ }\textbf {\bibinfo
  {volume} {8}},\ \bibinfo {pages} {122} (\bibinfo {year} {2006})},\ \Eprint
  {https://arxiv.org/abs/astro-ph/0606629} {arXiv:astro-ph/0606629 [astro-ph]}
  \BibitemShut {NoStop}%
\end{thebibliography}%

\end{document}